# Extended Probabilities

## Mathematical Foundations

**Mark Burgin**

Department of Mathematics
University of California, Los Angeles
405 Hilgard Ave.
Los Angeles, CA 90095

**Abstract**

There are important problems in physics related to the concept of probability. One of these problems is related to negative probabilities used in physics from 1930s. In spite of many demonstrations of usefulness of negative probabilities, physicists looked at them with suspicion trying to avoid this new concept in their theories. The main question that has bothered physicists is mathematical grounding and interpretation of negative probabilities. In this paper, we introduce extended probability as a probability function, which can take both positive and negative values. Defining extended probabilities in an axiomatic way, we show that classical probability is a positive section of extended probability.

*Keywords*: quantum mechanics, probability; negative probability; axiom; random experiment

## 1. Introduction

Contemporary physics and especially quantum physics essentially depend on mathematics they use. One of major mathematical concepts successfully employed quantum physics is probability. This concept gave birth to a specific mathematical theory called probability theory.

Probability theory is usually defined as the branch of mathematics concerned with analysis of random phenomena, for which probability is the central concept. However, there is no unique understanding of the term *probability* and there is no single mathematical system that formalizes the term *probability*, but rather a host of such systems. As a result, scientists who use probability techniques and statisticians have difficulties with grounded application of theoretical results. This causes many problems in probability theory and its applications. One of these problems is related to negative probabilities.

Negative probabilities emerged in physics in 1930s. History tells us that probability distributions were introduced previously by Dirac (1930) and Heisenberg (1931) within the context of quantum theory. However, both physicists missed its significance and possibility to take negative values, using this distribution as an approximation to the full quantum description of a system such as the atom. Wigner (1932) came to the conclusion that quantum corrections often lead to negative probabilities, or as some would say "quasi-probabilities". Wigner's goal was to supplant the wavefunction that appeared in Schrödinger's equation with a probability distribution in phase space. To do this, he introduced a function, which looked like a conventional probability distribution and has later been better known as the Wigner quasi-probability distribution because in contrast to conventional probability distributions, it took negative values, which could not be eliminated or made nonnegative. This clearly reveals the difference between classical and quantum probability as the Wigner quasi-probability distribution is the quantum analogue of the distribution function of classical statistical mechanics. The importance of Wigner's discovery for foundational problems was not recognized until much later.

In particular, Dirac (1942) not only supported Wigner's approach but also introduced the physical concept of negative energy. He wrote:

"*Negative energies and probabilities should not be considered as nonsense. They are well-defined concepts mathematically, like a negative of money.*"

After this, negative probabilities a little by little have become a popular although questionable technique in physics. Many physicists use negative probabilities (cf., for example, (Dirac, 1930; 1942; Heisenberg, 1931; Wigner, 1932; Khrennikov, 1993; 1995; Youssef, 1994; 1995; 2001; Scully, et al, 1994; Han, et al, 1996; Sokolovski, 2007; Bednorz and Belzig, 2009; Hofmann, 2009)). Feyman (1987) gave several examples demonstrating how negative probabilities naturally exist in physics and beyond.

Bartlett (1945) worked out the mathematical and logical consistency of negative probabilities. However, he did not establish rigorous foundation for negative probability utilization. In his book published in 2009, Khrennikov writes that it provides the first mathematical theory of negative probabilities. However, he is doing this not in the conventional setting of real numbers but in the framework of *p*-adic analysis. This is adequate not for the conventional physics in which the majority of physicists work but only for the so-called *p*-adic physics.

In this paper, we resolve the mathematical issue of the negative probability problem building a mathematical theory of extended probability as a probability function, which is defined for real numbers and can take both positive and negative values. As a result, extended probabilities include negative probabilities. Defining extended probabilities in an axiomatic way, we show that the classical probability is a positive section of extended probability.

**Constructions and denotation**

We remind that if $X$ is a set, then $|X|$ is the number of elements in (cardinality of) $X$ (Kuratowski and Mostowski, 1967). If $A \subseteq X$, then the complement of $A$ in $X$ is defined as $C_X A = X \setminus A$.

A system **B** of sets is called a *set ring* (Kolmogorov and Fomin, 1989) if it satisfies conditions (R1) and (R2):

(R1)  $A, B \in \mathbf{B}$ implies $A \cap B \in \mathbf{B}$.

(R2)  $A, B \in \mathbf{B}$ implies $A \Delta B \in \mathbf{B}$ where $A \Delta B = (A \setminus B) \cup (B \setminus A)$.

For any set ring $\mathbf{B}$, we have $\emptyset \in \mathbf{B}$ and $A, B \in \mathbf{B}$ implies $A \cup B, A \setminus B \in \mathbf{B}$.

Indeed, if $A \in \mathbf{B}$, then by R1, $A \setminus A = \emptyset \in \mathbf{B}$. If $A, B \in \mathbf{B}$, then $A \setminus B = ((A \setminus B) \cup (B \setminus A)) \cap A \in \mathbf{B}$. If $A, B \in \mathbf{B}$ and $A \cap B = \emptyset$, then $A \Delta B = A \cup B \in \mathbf{B}$. It implies that $A \cup B = (A \setminus B) \cup (B \setminus A) \cup (A \cap B) \in \mathbf{B}$. Thus, a system $\mathbf{B}$ of sets is a set ring if and only if it is closed with respect union, intersection and set difference.

The set **CI** of all closed intervals $[a, b]$ in the real line $\mathbf{R}$ is a set ring.

The set **OI** of all open intervals $(a, b)$ in the real line $\mathbf{R}$ is a set ring.

A set ring $\mathbf{B}$ with a unit element, i.e., an element $E$ from $\mathbf{B}$ such that for any $A$ from $\mathbf{B}$, we have $A \cap E = A$, is called a *set algebra* (Kolmogorov and Fomin, 1989).

The set **BCI** of all closed subintervals of the interval $[a, b]$ is a set algebra.

The set **BOI** of all open subintervals of the interval $[a, b]$ is a set algebra.

A set algebra $\mathbf{B}$ closed with respect to complement is called a *set field*.

2. **Mathematical concept of extended probability**

Extended probabilities generalize the standard definition of a probability function. At first, we define extended probabilities in an axiomatic way and then develop application of extended probabilities to finance.

To define extended probability, we need some concepts and constructions, which are described below.

Let us consider a set $\Omega$, which consists of two irreducible parts (subsets) $\Omega^+$ and $\Omega^-$, i.e., neither of these parts is equal to its proper subset, a set $\mathbf{F}$ of subsets of $\Omega$, and a function $P$ from $\mathbf{F}$ to the set $\mathbf{R}$ of real numbers.

Elements from $\mathbf{F}$, i.e., subsets of $\Omega$ that belong to $\mathbf{F}$, are called *random events*.

Elements from $\mathbf{F}^+ = \{X \in \mathbf{F}; X \subseteq \Omega^+\}$ are called *positive random events*.

Elements from $\Omega^+$ that belong to $\mathbf{F}^+$ are called *elementary positive random events* or simply, *elementary positive random events*.

If $w \in \Omega^+$, then $-w$ is called the *antievent* of $w$.

Elements from $\Omega^-$ that belong to $\mathbf{F}^-$ are called *elementary negative random events* or *elementary random antievents*.

For any set $X \subseteq \Omega^+$, we define

$$X^+ = X \cap \Omega^+,$$
$$X^- = X \cap \Omega^-,$$
$$-X = \{-w; w \in X\}$$

and

$$\mathbf{F}^- = \{-A; A \in \mathbf{F}^+\}$$

If $A \in \mathbf{F}^+$, then $-A$ is called the *antievent* of $A$.

Elements from $\mathbf{F}^-$ are called *negative random events* or *random antievents*.

**Definition 1.** The function $P$ from $\mathbf{F}$ to the set $\mathbf{R}$ of real numbers is called a *probability function*, if it satisfies the following axioms:

**EP 1** (*Order structure*). There is a graded involution $\alpha: \Omega \to \Omega$, i.e., a mapping such that $\alpha^2$ is an identity mapping on $\Omega$ with the following properties: $\alpha(w) = -w$ for any element $w$ from $\Omega$, $\alpha(\Omega^+) \supseteq \Omega^-$, and if $w \in \Omega^+$, then $\alpha(w) \notin \Omega^+$.

**EP 2** (*Algebraic structure*). $\mathbf{F}^+ \equiv \{X \in \mathbf{F}; X \subseteq \Omega^+\}$ is a set algebra that has $\Omega^+$ as a member.

**EP 3** (*Normalization*). $P(\Omega^+) = 1$.

**EP 4** (*Composition*) $\mathbf{F} \equiv \{X; X^+ \subseteq \mathbf{F}^+ \ \& \ X^- \subseteq \mathbf{F}^- \ \& \ X^+ \cap -X^- \equiv \emptyset \ \& \ X^- \cap -X^+ \equiv \emptyset\}$.

**EP 5** (*Finite additivity*)
$$P(A \cup B) = P(A) + P(B)$$
for all sets $A, B \in \mathbf{F}$ such that
$$A \cap B \equiv \emptyset$$

**EP 6** (*Annihilation*). $\{v_i, w, -w; v_i, w \in \Omega \ \& \ i \in I\} = \{v_i; v_i \in \Omega \ \& \ i \in I\}$ for any element $w$ from $\Omega$.

Axiom EP6 shows that if $w$ and $-w$ are taken (come) into one set, they annihilate one another. Having this in mind, we use two equality symbols: $=$ and $\equiv$. The second symbol means equality of elements of sets. The second symbol also means equality of sets, when two sets are equal when they have exactly the same elements (Kuratowski and Mostowski, 1967). The equality symbol $=$ is used to denote equality of two sets with annihilation, for example, $\{w, -w\} = \emptyset$. Note that for sets, equality $\equiv$ implies equality $=$.

For equality of numbers, we, as it is customary, use symbol $=$.

**EP 7**. (*Adequacy*) $A = B$ implies $P(A) = P(B)$ for all sets $A, B \in \mathbf{F}$.

For instance, $P(\{w, -w\}) = P(\emptyset) = 0$.

**EP 8**. (*Non-negativity*) $P(A) \geq 0$, for all $A \in \mathbf{F}^+$.

**Remark 1**. It is known that for any set algebra $\mathbf{A}$, the empty set $\emptyset$ belongs to $\mathbf{A}$ and for any set field $\mathbf{B}$ in $\Omega$, the set $\Omega$ belongs to $\mathbf{A}$ (Kolmogorov and Fomin, 1989).

**Definition 2.** The triad $(\Omega, \mathbf{F}, P)$ is called an *extended probability space*.

**Definition 3.** If $A \in \mathbf{F}$, then the number $P(A)$ is called the *extended probability* of the event $A$.

Let us obtain some properties of the introduced constructions.

**Lemma 1.** $\alpha(\Omega^+) \equiv -\Omega^+ \equiv \Omega^-$ and $\alpha(\Omega^-) \equiv -\Omega^- \equiv \Omega^+$.

Proof. By Axiom EP1, $\alpha(\Omega^+) \equiv -\Omega^+$ and $\alpha(\Omega^+) \supseteq \Omega^-$. As $\Omega \equiv \Omega^+ \cup \Omega^-$, Axiom EP1 also implies $\alpha(\Omega^+) \subseteq \Omega^-$. Thus, we have $\alpha(\Omega^+) \equiv \Omega^-$. The first part is proved.

The second part is proved in a similar way.

Thus, if $\Omega^+ = \{w_i; i \in I\}$, then $\Omega^- = \{-w_i; i \in I\}$.

As $\alpha$ is an involution of the whole space, we have the following result.

**Proposition 1.** $\alpha$ is a one-to-one mapping and $|\Omega^+| = |\Omega^-|$.

**Corollary 1**. (*Domain symmetry*) $w \in \Omega^+$ if and only if $-w \in \Omega^-$.

**Corollary 2**. (*Element symmetry*) $-(-w) = w$ for any element $w$ from $\Omega$.

**Corollary 3**. (*Event symmetry*) $-(-X) \equiv X$ for any event $X$ from $\Omega$.

**Lemma 2.** $\alpha(w) \neq w$ for any element $w$ from $\Omega$.

Indeed, this is true because if $w \in \Omega^+$, then by Axiom EP1, $\alpha(w) \notin \Omega^+$ and thus, $\alpha(w) \neq w$. If $w \in \Omega^-$, then we may assume that $\alpha(w) = w$. However, in this case, $\alpha(v) = w$ for some element $v$ from $\Omega^+$ because by Axiom EP1, $\alpha$ is a projection of $\Omega^+$ onto $\Omega^-$. Consequently, we have

$$\alpha(\alpha(v)) = \alpha(w) = w$$

However, $\alpha$ is an involution, and we have $\alpha(\alpha(v)) = v$. This results in the equality

$$v = w$$

. Consequently, we have $\alpha(v) = v$. This contradicts Axiom EP1 because $v \in \Omega^+$. Thus, lemma is proved by contradiction.

**Proposition 2.** $\Omega^+ \cap \Omega^- \equiv \emptyset$.

<u>Proof</u>. Let $w \in \Omega^+ \cap \Omega^-$. Then by Axiom EP1, $-w \in \Omega \setminus \Omega^+ = \Omega^- \setminus \Omega^+$ as $\Omega = \Omega^- \cup \Omega^+$ By Axiom EP6, $\Omega^- = \Omega^- \setminus \{w\}$. However, this contradicts irreducibility of $\Omega^-$.

Proposition is proved by contradiction.

**Proposition 3.** $\mathbf{F}^+ \subseteq \mathbf{F}$, $\mathbf{F}^- \subseteq \mathbf{F}$ and $\mathbf{F} \subseteq \mathbf{F}^+ \cup \mathbf{F}^-$.

Indeed, if $X \subseteq \Omega^+$, then $X = X^+$. If $X \in \mathbf{F}^+$, then by Axiom EP2, $X \in \mathbf{F}$ because $X^- = \emptyset$ and $\emptyset \in \mathbf{F}^-$ (cf. Remark 1). Thus, $\mathbf{F}^+ \subseteq \mathbf{F}$ as the set $X$ is an arbitrary element from $\mathbf{F}^+$.

By the same token, $\mathbf{F}^- \subseteq \mathbf{F}$.

By Axiom EP4, $\mathbf{F} \subseteq \mathbf{F}^+ \cup \mathbf{F}^-$.

Proposition is proved.

Corollary 1 implies the following result.

**Proposition 4.** $X \subseteq \Omega^+$ if and only if $- X \subseteq \Omega^-$.

**Proposition 5.** $\mathbf{F}^- \equiv \{X \in \mathbf{F}; X \subseteq \Omega^-\} = \mathbf{F} \cap \Omega^-$.

**Corollary 4.** $\mathbf{F}^+ \cap \mathbf{F}^- \equiv \emptyset$.

Axioms EP6 implies the following result.

**Lemma 3.** $X \cup -X = \emptyset$ for any subset $X$ of $\Omega$.

Indeed, for any $w$ from the set $X$, there is $-w$ in the set $X$, which annihilates $w$.

Let us define the union with annihilation of two subsets $X$ and $Y$ of $\Omega$ by the following formula:

$$X + Y \equiv (X \cup Y) \setminus [(X \cap -Y) \cup (-X \cap Y)]$$

Here the set-theoretical operation $\setminus$ represents annihilation, while sets $X \cap -Y$ and $X \cap -Y$ depict annihilating entities.

Some properties of the new set operation $+$ are the same as properties of the union $\cup$, while other properties are different. For instance, there is no distributivity between operations $+$ and $\cap$.

**Lemma 4.** a) $X + X \equiv X$ for any subset $X$ of $\Omega$;

b) $X + Y \equiv X + Y$ for any subsets $X$ and $Y$ of $\Omega$;

c) $X + \emptyset \equiv X$ for any subset $X$ of $\Omega$;

d) $X + (Y + Z) \equiv (X + Y) + Z$ for any subsets $X$, $Y$ and $Z$ of $\Omega$;

e) $X + Y \equiv X \cup Y$ for any subsets $X$ and $Y$ of $\Omega^+$ (of $\Omega^-$);

**Lemma 5.** a) $Z \cap (X + Y) \neq Z \cap X + Z \cap Y$;

b) $X + (Y \cap Z) \neq (X \cap Y) + (X \cap Z)$.

**Lemma 6.** $A \cap B \equiv (A^+ \cap B^+) + (A^- \cap B^-)$ for any subsets $A$ and $B$ of $\Omega$.

Indeed, as $A \equiv A^+ \cup A^-$ and $B \equiv B^+ \cup B^-$, we have

$$A \cap B \equiv (A^+ \cup A^-) \cap (B^+ \cap B^-) \equiv$$

$$(A^+ \cap B^+) \cup (A^+ \cap B^-) \cup (A^- \cap B^+) \cup (A^- \cap B^-) \equiv$$

$$(A^+ \cap B^+) + (A^- \cap B^-)$$

because $(A^+ \cap B^-) \equiv \emptyset$ and $(A^- \cap B^+) \equiv \emptyset$.

In a similar way, we prove the following result.

**Lemma 7.** $A \setminus B \equiv (A^+ \setminus B^+) + (A^- \setminus B^-)$ for any subsets $A$ and $B$ of $\Omega$.

**Lemma 8.** $X \equiv X^+ + X^- = X^+ \cup X^-$ for any set $X$ from **F**.

Indeed, as $X \subseteq \Omega^+$, $X^+ = X \cap \Omega^+$, $X^- = X \cap \Omega^-$, and $\Omega = \Omega^+ \cup \Omega^-$, we have $X = X^+ \cup X^-$ as $\Omega \equiv \Omega^+ \cup \Omega^-$. In addition, $X^+ + X^- = X^+ \cup X^-$ because by Axiom EP 4, we have $X^+ \cap -X^- = \emptyset$  $X^- \cap -X^+ = \emptyset$.

**Lemma 9.** $A + B \equiv (A^+ + B^+) + (A^- + B^-)$ for any sets $X$ and $Y$ from **F**.

Indeed, as by Lemma 8, $A \equiv A^+ + A^-$ and $B \equiv B^+ + B^-$, we have

$$A + B \equiv (A^+ + A^-) + (B^+ + B^-) \equiv$$

$$(A^+ + B^+) + (A^- + B^-)$$

because by Lemma 5, operation + is commutative and associative.

**Lemma 10.** $P(\emptyset) = 0$.

Indeed, by Axiom EP5, we have $P(A \cup \emptyset) = P(A) + P(\emptyset)$. Thus, $P(\emptyset) = P(A \cup \emptyset) - P(A) = P(A) - P(A) = 0$.

Lemma 10 has the following interpretation. In each trial (experiment) something happens. So, the extended probability that nothing happens is equal to zero.

Proposition 2 implies the following result.

**Lemma 11.** $X^+ \cap X^- = \emptyset$ for any $X \subseteq \Omega$.

Properties of the structure $\mathbf{F}^+$ are mirrored in the structure $\mathbf{F}^-$.

**Theorem 1.** (*Algebra symmetry*) If $\mathbf{F}^+$ is a set algebra (or set field), then $\mathbf{F}^-$ is a set algebra (or set field).

<u>Proof</u>. Let us assume that $\mathbf{F}^+$ is a set algebra and take two negative random events $H$ and $K$ from $\mathbf{F}^-$. By the definition of $\mathbf{F}^-$, $H = -A$ and $K = -B$ for some positive random events $A$ and $B$ from $\mathbf{F}^+$. Then we have

$$H \cap K = (-A) \cap (-B) = -(A \cap B)$$

As $\mathbf{F}^+$ is a set algebra, $A \cap B \in \mathbf{F}^+$. Thus, $H \cap K \in \mathbf{F}^-$.

In a similar way, we have

$$H \cup K = (-A) \cup (-B) = -(A \cup B)$$

As $\mathbf{F}^+$ is a set algebra, $A \cup B \in \mathbf{F}^+$. Thus, $H \cup K \in \mathbf{F}^-$.

By the same token, we have $H \setminus K \in \mathbf{F}^-$.

Besides, if $\mathbf{F}^+$ has a unit element $E$, then $-E$ is a unit element in $\mathbf{F}^-$.

Thus, $\mathbf{F}^-$ is a set algebra.

Now let us assume that $\mathbf{F}^+$ is a set field and $H \in \mathbf{F}^-$. Then by the definition of $\mathbf{F}^-$, $H = -A$ for a positive random event $A$ from $\mathbf{F}^+$. It means that $C_{\Omega^+}A = \Omega^+ \setminus A \in \mathbf{F}^+$. At the same time,

$$C_{\Omega^-}H = \Omega^- \setminus H = (-\Omega^+) \setminus (-A) = -(\Omega^+ \setminus A) = -C_{\Omega^+}A$$

As $C_{\Omega^+}A$ belongs to $\mathbf{F}^+$, the complement $C_{\Omega^-}H$ of $H$ belongs to $\mathbf{F}^-$. Consequently, $\mathbf{F}^-$ is a set field.

Theorem is proved.

Axioms EP6 and EP7 imply the following result.

**Proposition 6.** $P(X + Y) = P(X \cup Y)$ for any two random events $X$ and $Y$ from $\mathbf{F}$.

Properties of the structure $\mathbf{F}^+$ are inherited by the structure $\mathbf{F}$.

**Theorem 2.** If $\mathbf{F}^+$ is a set field (or set algebra), then $\mathbf{F}$ is a set field (or set algebra) with respect to operations $+$ and $\cap$.

Proof. Let us assume that $\mathbf{F}^+$ is a set algebra and take two random events $A$ and $B$ from $\mathbf{F}$. Then by Theorem 1, $\mathbf{F}^-$ is a set algebra. By Lemma 8, $A \equiv A^+ + A^-$ and $B \equiv B^+ + B^-$. By Axiom EP4, $A^+, B^+ \in \mathbf{F}^+$, $A^-, B^- \in \mathbf{F}^-$, while by Proposition 2, $A^+ \cap A^- \equiv \emptyset$, $B^+ \cap B^- \equiv \emptyset$, $A \equiv A^+ \cup A^-$, and $B \equiv B^+ \cup B^-$.

By Lemma 6, $A \cap B \equiv (A^+ \cap B^+) + (A^- \cap B^-)$. Thus, $(A \cap B)^+ \equiv A^+ \cap B^+$ and $(A \cap B)^- \equiv A^- \cap B^-$. As $\mathbf{F}^+$ is a set algebra, $(A \cap B)^+ \equiv A^+ \cap B^+ \in \mathbf{F}^+$. As by Theorem 1, $\mathbf{F}^-$ is a set algebra, $(A \cap B)^- \equiv A^- \cap B^- \in \mathbf{F}^-$. Consequently, $A \cap B \in \mathbf{F}$.

By Lemma 7, $A \setminus B \equiv (A^+ \setminus B^+) + (A^- \setminus B^-)$. Thus, $(A \setminus B)^+ \equiv A^+ \setminus B^+$ and $(A \setminus B)^- \equiv A^- \setminus B^-$. As $\mathbf{F}^+$ is a set algebra, $(A \setminus B)^+ \equiv A^+ \setminus B^+ \in \mathbf{F}^+$. As by Theorem 1, $\mathbf{F}^-$ is a set algebra, $(A \setminus B)^- \equiv A^- \setminus B^- \in \mathbf{F}^-$. Consequently, $A \setminus B \in \mathbf{F}$.

By Lemma 9, $A + B \equiv (A^+ + B^+) + (A^- + B^-)$. Thus, $(A + B)^+ \equiv A^+ + B^+$ and $(A + B)^- \equiv A^- + B^-$. As $\mathbf{F}^+$ is a set algebra, $(A + B)^+ \equiv A^+ + B^+ \equiv A^+ \cup B^+ \in \mathbf{F}^+$. As by Theorem 1, $\mathbf{F}^-$ is a set algebra, $(A + B)^- \equiv A^- + B^- \equiv A^- \cup B^- \in \mathbf{F}^-$. Consequently, $A + B \in \mathbf{F}$.

Besides, if $\mathbf{F}^+$ has a unit element $E$, then $-E$ is a unit element in $\mathbf{F}^-$ and $E \cup -E$ is a unit element in $\mathbf{F}$.

Thus, $\mathbf{F}$ is a set algebra.

Now let us assume that $\mathbf{F}^+$ is a set field and $A \in \mathbf{F}$. Then by Theorem 1, $\mathbf{F}^-$ is a set field. By Lemma 8, $A \equiv A^+ + A^-$. By Proposition 2, $\Omega^+ \cap \Omega^- = \emptyset$, we have

$$C_\Omega A = C_{\Omega^+} A^+ + C_{\Omega^-} A^-$$

Then $C_{\Omega^+} A^+$ belongs to $\mathbf{F}^+$ as $\mathbf{F}^+$ is a set field, while as it is proved in Theorem 1, $C_{\Omega^-} A^-$ belongs to $\mathbf{F}^-$. Consequently, $C_\Omega A$ belongs to $\mathbf{F}$ and $\mathbf{F}$ is a set field.

Theorem is proved.

**Proposition 7.** $X \cap -Y = -(-X \cap Y)$.

Indeed, if $w \in X$, then $-w \in -X$. At the same time, $w \in -Y$, then $-w \in -(-Y) = Y$ by Corollary 3. Thus, $-w \in -X \cap Y$ and $w \in -(-X \cap Y)$. By the same token, if $w \in -(-X \cap Y)$, then $w \in -(-X \cap Y)$.

**Proposition 8.** $P(A) = - P(-A)$ for any random event $A$ from **F**.

Proof. By Lemma 3, $A \cup -A = \emptyset$. By Axiom EP6, $P(A \cup -A) = P(\emptyset)$. By Axiom EP5, $P(A \cup -A) = P(A) + P(-A)$ as $A \cap -A = \emptyset$. By Lemma 10, $P(\emptyset) = 0$. Thus, $P(A) + P(-A) = 0$ and $P(A) = - P(-A)$ for any random event $A$ from **F**.

**Corollary 5.** (*Non-positivity*) $P(A) \leq 0$, for all $A \in \mathbf{F}^-$.

The extended probability of a random event is composed from two components as the following theorem shows.

**Theorem 3.** $P(A) = P(A^+) - P(-A^-) = P(A^+) + P(A^-)$ for any random event $A \in \mathbf{F}$.

Proof. By Axiom EP 4, $A = A^+ \cup A^-$. Then by Axiom EP 5, $P(A) = P(A^+) + P(-A^-)$ because by Lemma 11, $X^+ \cap X^- = \emptyset$ for any subset $X$ of $\Omega$. By Proposition 8, $P(A^-) = - P(-A^-)$. Consequently, $P(A) = P(A^+) - P(-A^-)$.

A similar result is valid for complements of random events.

**Theorem 4.** $P(C_\Omega A) = P(C_{\Omega+} A^+) - P(C_{\Omega+}(-A^-)) = P(C_{\Omega+} A^+) - P(-C_\Omega A^-)$ for any random event $A \in \mathbf{F}$.

Proof. As it is demonstrated in the proof of Theorem 2, $C_\Omega A = C_{\Omega+} A^+ + C_\Omega A^-$. Consequently, $(C_\Omega A)^+ = C_{\Omega+} A^+$ and $(C_\Omega A)^- = C_{\Omega+} A^-$. As $C_{\Omega+} A^- = -C_{\Omega+}(-A^-) = -(-C_\Omega A^-)$, Theorem 4 follows from Theorem 3.

**Proposition 9.** $P(\Omega) = 0$.

Proof. By Proposition 4, $P(\Omega) = P(\Omega^+ \cup \emptyset) = P(\Omega^+) + P(\emptyset) = P(\Omega^+)$. Thus, $P(\emptyset) = P(\Omega^+) - P(\Omega^+) = 0$.

Proposition 9 has the following interpretation. The set $\Omega$ cannot exist (be stable) due to annihilation. So, its extended probability, or more adequately, the extended probability of its existence, is equal to zero.

There is a relation between the extended probability of a random event and the extended probability of the complement of this random event. However, this relation is very different from the relation that exists between the probability of a random event and the probability of the complement of this random event. For probabilities, we have the following relation (cf., for example (Kolmogorov, 1933):

$$p(C_\Omega A) = 1 - p(A)$$

**Proposition 10.** $P(A) = - P(C_\Omega A)$ for all $A \in \mathbf{F}$.

Indeed, taking an arbitrary set $A$ from $\mathbf{F}$, $A \cup C_\Omega A = \Omega$. Thus, by Axiom EP5, $P(A) + P(C_\Omega A) = P(A \cup C_\Omega A) = P(\Omega) = 0$. Consequently, $P(A) = - P(C_\Omega A)$.

**Proposition 11**. If $A = \{ w_1, w_2, w_3, \ldots, w_k \}$ belongs to $\mathbf{F}$ and all $w_1, w_2, w_3, \ldots, w_k$ belongs to $\mathbf{F}$, then $P(A) = P(w_1) + P(w_2) + P(w_3) + \ldots + P(w_k)$.

Proposition 11 directly follows from Axiom EP5 because $A$ is a set of elementary random events.

**Theorem 4**. Any probability function $P$ is monotone on $\mathbf{F}^+$, i.e., if $A \subseteq B$ and $A, B \in \mathbf{F}^+$, then $P(A) \leq P(B)$, and is antimonotone on $\mathbf{F}^-$, i.e., if $H \subseteq K$ and $H, K \in \mathbf{F}^-$, then $P(H) \geq P(K)$.

Proof. Let us consider two random events $A$ and $B$ from $\mathbf{F}^+$, such that $A \subseteq B$. In this case, $B = A \cup C$ for some random event $C$ from $\mathbf{F}^+$ where $A \cap C = \emptyset$. By Axiom EP5, $P(B) = P(A) + P(C)$ and by Axiom EP8, both $P(A)$ and $P(C)$ are non-negative numbers. Thus, $P(A) \leq P(B)$.

Let us consider two random events $H$ and $K$ from $\mathbf{F}^-$, such that $H \subseteq K$. In this case, $K = H \cup G$ for some random event $G$ from $\mathbf{F}^-$ where $H \cap G = \emptyset$. By Axiom EP5, $P(K) = P(H) + P(G)$ and by Corollary 5, both $P(H)$ and $P(G)$ are not positive numbers. Thus, $P(K) \leq P(H)$.

Theorem is proved.

**Proposition 11**. $1 \geq P(A) \geq -1$, for all $A \in \mathbf{F}$.

Proof. By Axiom EP3, $P(\Omega^+) = 1$. Then by Proposition 8, $P(\Omega^-) = -1$. By Theorem 4, $P(A) \leq 1$ for any random event $A$ from $\mathbf{F}^+$ and $P(B) \geq -1$ for any random event $B$ from $\mathbf{F}^-$. By Theorem 3, $P(A) = P(A^+) + P(A^-)$ for any random event $A \in \mathbf{F}$. Thus, $P(A)$ cannot be larger than 1 because both $P(A^+)$ and $P(A^-)$ have absolute values less than 1 and they have opposite signs.

Proposition is proved.

As in the case of the classical probability, it is possible to add one more axiom called *axiom of continuity* to the list of axioms for extended probability. This allows us to comply with the traditional approach to probability.

**EP 9**. (*Continuity*) If

$$A_1 \supseteq A_2 \supseteq A_3 \supseteq \ldots \supseteq A_i \supseteq \ldots$$

is a decreasing sequence of events $A_i$ from $\mathbf{F}^+$ such that

$$\bigcap_{i=1}^{\infty} A_i = \emptyset,$$

then

$$\lim_{i \to \infty} P(A_i) = 0.$$

**Theorem 5**. All events from the algebra $\mathbf{F}$ have the continuity property.

Proof. Let us take a decreasing sequence of events $A_i$ from $\mathbf{F}$

$$A_1 \supseteq A_2 \supseteq A_3 \supseteq \ldots \supseteq A_i \supseteq \ldots$$

such that

$$\bigcap_{i=1}^{\infty} A_i = \emptyset$$

By Proposition 10, any subset $X$ of the set $\Omega$ has the form $X = X^+ \cup X^-$. Thus, $A_i = A_i^+ \cup A_i^-$ for all $i = 1, 2, 3, \ldots$ As by Proposition 9, $X^+ \cap X^- = \emptyset$ for any subset $X$ of $\Omega$, we have two decreasing sequences of events

$$A_1^+ \supseteq A_2^+ \supseteq A_3^+ \supseteq \ldots \supseteq A_i^+ \supseteq \ldots$$

and

$$A_1^- \supseteq A_2^- \supseteq A_3^- \supseteq \ldots \supseteq A_i^- \supseteq \ldots$$

As $A_i \supseteq A_i^+$ for all $i = 1, 2, 3, \ldots$ , we have

$$\bigcap_{i=1}^{\infty} A_i^+ = \emptyset$$

as all events $A_i^+$ belong to $\mathbf{F}^+$, Axiom EP9 implies

$$\lim_{i \to \infty} P(A_i^+) = 0.$$

Besides, $A_i \supseteq A_i^-$ for all $i = 1, 2, 3, \ldots$ , and thus,

$$\bigcap_{i=1}^{\infty} A_i^- = \emptyset$$

Consequently,

$$\bigcap_{i=1}^{\infty} -A_i^- = \emptyset$$

and by definition, all events $-A_i^-$ belong to $\mathbf{F}^+$

Consequently, by Axiom EP9, we have

$$\lim_{i \to \infty} P(-A_i^-) = 0.$$

By Axiom EP 6, $P(A_i^-) = -P(-A_i^-)$. So, we have

$$\lim_{i \to \infty} P(A_i^-) = 0.$$

By Theorem 3, $P(A_i) = P(A_i^+) + P(A_i^-)$ for all $i = 1, 2, 3, \ldots$ and by properties of limits, we have

$$\lim_{i \to \infty} P(A_i) = \lim_{i \to \infty} P(A_i^+) + \lim_{i \to \infty} P(A_i^-) = 0 + 0 = 0.$$

Theorem is proved as the chosen sequence is arbitrary.

**EP 10** (*Decomposition*) For any $A \in \mathbf{F}$, we have
$$P(A) = P(A^+) + P(A^-)$$

It is possible to consider a restricted form of Axiom EP5.

**EP 5p** (*Positive finite additivity*)
$$P(A \cup B) = P(A) + P(B)$$

for all sets $A, B \in \mathbf{F}^+$ such that

$$A \cap B = \emptyset \text{ and } A \cap -B = \emptyset.$$

**Theorem 6**. Axiom EP5p and EP10 imply Axiom EP5.

Problems with probability interpretations and necessity to have sound mathematical foundations brought forth an axiomatic approach in probability theory. Based on ideas of Fréchet and following the axiomatic mainstream in mathematics, Kolmogorov developed his famous axiomatic exposition of probability theory (1933). Here we show that extended probabilities really extend the classical concept of probability axiomatized by Kolmogorov.

**Theorem 7**. Restricted to the positive set algebra $\mathbf{F}^+$, an extended probability function $P(A)$ is a conventional probability function, which satisfies Kolmogorov's axioms for probability.

Indeed, if we restrict the probability function $P(A)$ to the positive set algebra (set field) $\mathbf{F}^+ = \{X \in \mathbf{F}; X \subseteq \Omega^+ \}$, we will have the following axioms.

**EP 2** (*Algebraic structure*). $\mathbf{F}^+ = \{X \in \mathbf{F}; X \subseteq \Omega^+ \}$ is a set field (or set algebra), i. e., a set of subsets of $\Omega^+$ that has $\Omega^+$ as a member, and that is closed under complementation (with respect to $\Omega^+$) and union.

**EP 3** (*Normalization*). $P(\Omega^+) = 1$.

**EP 5** (*Finite additivity*)

$$P(A \cup B) = P(A) + P(B)$$

for all sets $A, B \in \mathbf{F}^+$ such that

$$A \cap B = \emptyset.$$

**EP 8**. (*Non-negativity*) $P(A) \geq 0$, for all $A \in \mathbf{F}^+$.

The first axiom defines the algebraic structure used for classical probability, while other three axioms coincide with Kolmogorov's axioms for probability (Kolmogorov, 1933):

**K 1**. (*Non-negativity*) $P(A) \geq 0$, for all $A \in \mathbf{F}$.

**K 2**. (*Normalization*) $P(\Omega) = 1$.

**K 3**. (*Finite additivity*)

$$P(A \cup B) = P(A) + P(B)$$

for all sets $A, B \in \mathbf{F}$ such that

$$A \cap B = \emptyset.$$

4.  **Conclusion**

We have developed a mathematical theory of extended probabilities, which include negative probabilities, giving axioms for extended probabilities and deriving their general properties. The concept of extended probability eliminates many problems that researchers have with negative probabilities.

**References**

1. Allen, E.H. Negative Probabilities and the Uses of Signed Probability Theory, *Philosophy of Science*, Vol. 43, No. 1 (1976), pp. 53-70
2. Bartlett, M. S. (1945) Negative Probability, *Math Proc Camb Phil Soc* **41**: 71–73
3. Bednorz, A. and Belzig, W. On the problem of negative probabilities in time-resolved full counting statistics, 25th International Conference on Low Temperature Physics (LT25), *Journal of Physics*, Conference Series, v. 150 (2009) 022005
4. Dirac, P.A.M. (1930) Note on exchange phenomena in the Thomas atom, *Proc. Camb. Phil. Soc.* v. 26, pp. 376-395
5. Dirac, P.A.M. (1942) The Physical Interpretation of Quantum Mechanics, *Proc. Roy. Soc. London*, (A 180), pp. 1–39
6. Feynman, R. P. The Concept of Probability Theory in Quantum Mechanics, in the *Second Berkeley Symposium on Mathematical Statistics and Probability Theory*, University of California Press, Berkeley, California, 1950
7. Feynman, R. P. (1987) Negative Probability, in *Quantum Implications*: *Essays in Honour of David Bohm*, Routledge & Kegan Paul Ltd, London & New York, pp. 235–248
8. Han, Y.D., Hwang, W.Y. and Koh, I.G. Explicit solutions for negative-probability measures for all entangled states, *Physics Letters* A, v. 221, No. 5, 1996, pp. 283-286
9. Hartle, J. Quantum mechanics with extended probabilities, *Phys. Rev.* A 78, (2008).
10. Heisenberg, W. (1931) Über die inkohärente Streuung von Röntgenstrahlen, *Physik. Zeitschr.* v. 32, pp. 737-740

11. Hofmann, H.F. (2009) How to simulate a universal quantum computer using negative probabilities, *J. Phys.* A: Math. Theor. 42 275304 (9pp)

12. Hotson, D. Dirac's Equation and the Sea of Negative Energy, *Infinite Energy*, No. 43/44, 2002

13. Khrennikov, A. Yu. (1993) *p*-adic theory of probability and its applications. A principle of the statistical stabilization of frequencies,*Teoreticheskaia i Matematicheskaia Fizika*, v. 97(3), pp. 348–363 (in Russian)

14. Khrennikov, A. *p-adic probability distributions of hidden variables*, SFB 237 Preprint 257, 1995.

15. Khrennikov, A. *p-adic probability interpretation of Bell's inequality*, Ph. Lett. A 200 (1995).

16. Khrennikov, A. *Interpretations of Probability*, Walter de Gruyter, Berlin/New York, 2009

17. Kolmogorov, A. N. (1933) *Grundbegriffe der Wahrscheinlichkeitrechnung*, Ergebnisse Der Mathematik (English translation: (1950) *Foundations of the Theory of Probability*, Chelsea P.C.)

18. Kolmogorov, A.N. and Fomin, S.V. *Elements of Function Theory and Functional Analysis*, Nauka, Moscow, 1989    (in Russian)

19. Kuratowski, K. and Mostowski, A. *Set Theory*, North Holland P.C., Amsterdam, 1967

20. Mardari, G.N. (2007) Interpreting Negative Probabilities in the Context of Double-Slit Interferometry, in 4th *Conference on Foundations of Probability and Physics*, Växjö, Sweden, pp.341-346

21. Scully, M.O., Walther, H.and Schleich, W. *Feynman's approach to negative probability in quantum mechanics*, Phys.Rev. A49, 1562(1994)

22. Sokolovski, D. (2007) Weak values, "negative probability," and the uncertainty principle, *Phys. Rev.* A 76, 042125

23. Streater, R.F. *Classical and Quantum Probability*, Preprint math-ph/0002049, 2000 (electronic edition:   http://arXiv.org)

24. Wigner, E.P. (1932) On the quantum correction for thermodynamic equilibrium, *Phys. Rev.* v. 40, pp. 749-759

25. Youssef, S. Quantum Mechanics as Complex Probability Theory, *Mod.Phys.Lett.* A9, 2571 (1994)

26. Youssef, S.Is Quantum Mechanics an Exotic Probability Theory?, in Fundamental Problems in Quantum Theory, Conference in Honor of Professor John A. Wheeler, *Annals of the New York Academy of Sciences*, v. 755, April, 1995

27. Youssef, S. Quantum Mechanics as an Exotic Probability Theory, in *Proceedings of the Fifteenth International Workshop on Maximum Entropy and Bayesian Methods*, Santa Fe, August, 1995

28. Youssef, S. (1995) Is Complex Probability Theory Consistent with Bell's Theorem?, *Phys.Lett.* A204, 181
29. Youssef, S. *Physics with exotic probability theory*, Preprint hep-th/0110253, 2001 (electronic edition:   http://arXiv.org)